\documentclass[preprint,twocolumn,proceedings,newstyle]{rmaa}

\newcommand{\D}{\discretionary{}{}{}}

\SetYear{2002}
\SetConfTitle{Galaxy Evolution: Theory and Observations}

\title{SAURON: Observations of E/S0/Sa galaxies}

\author{H.~Kuntschner\altaffilmark{1}, R.~Bacon\altaffilmark{2},
  M.~Bureau\altaffilmark{3}, M.~Cappellari\altaffilmark{4},
  Y.~Copin\altaffilmark{5}, R.~L.~Davies\altaffilmark{6},
  E.~Emsellem\altaffilmark{2}, B.~W.~Miller\altaffilmark{7},
  R.~McDermid\altaffilmark{6}, R.~F.~Peletier\altaffilmark{2,8},
  E.~K.~Verolme\altaffilmark{4} and P.~T.~de~Zeeuw\altaffilmark{4}}
  
\altaffiltext{1}{European Southern Observatory, Karl-Schwarzschild-Str.
  2, 85748 Garching, Germany (hkuntsch\D{}@eso.\D{}org).}
\altaffiltext{2}{Observatoire de Lyon, France.}
\altaffiltext{3}{Columbia University, US.}
\altaffiltext{4}{Sterrewacht Leiden, Netherlands.}
\altaffiltext{5}{Institut de Physique Nucleaire de Lyon, France.}
\altaffiltext{6}{University of Durham, UK.}
\altaffiltext{7}{Gemini Observatory, Chile.}
\altaffiltext{8}{University of Nottingham, UK.}

\suppressfulladdresses

\listofauthors{H.~Kuntschner, R.~Bacon, M.~Bureau, M.~Cappellari,
    Y.~Copin, R.~L.~Davies, E.~Emsellem, B.~W.~Miller, R.~McDermid,
    R.~F.~Peletier, E.~K.~Verolme, \& P.~T.~de~Zeeuw}
\indexauthor{Kuntschner, H.}
\indexauthor{Bacon, R.} 
\indexauthor{Bureau, M.} 
\indexauthor{Cappellari, M.}
\indexauthor{Copin, Y.}
\indexauthor{Davies, R. L.}  
\indexauthor{Emsellem, E.} 
\indexauthor{Miller, B. W.} 
\indexauthor{McDermid, R.}
\indexauthor{Peletier, R. F.}  
\indexauthor{Verolme, E. K.} 
\indexauthor{de Zeeuw, P. T.}

\addkeyword{galaxies: elliptical and lenticular}
\addkeyword{galaxies: formation}
\addkeyword{galaxies: kinematics and dynamics}
\addkeyword{galaxies: fundamental parameters}
\addkeyword{surveys}

\begin{document}
\maketitle 

\boldabstract{We present results from a new and unique integral-field
  spectrograph, SAURON.} It has a large field of view and high
throughput and is primarily built for the study of stellar \& gaseous
kinematics and stellar populations in galaxies. Its aim is to carry out
a systematic survey of the velocity fields, velocity dispersions, and
line-strength distributions of nearby ellipticals, lenticular galaxies
and spiral bulges. Its wide field is especially useful for the study of
complicated velocity structures. Together with other spectroscopic
data, images, and dynamical modelling, SAURON will help to constrain the
intrinsic shapes, mass-to-light ratios, and stellar populations of
early-type galaxies and spiral bulges.


SAURON (Spectroscopic Areal Unit for Research on Optical Nebulae) is a
TIGER-like integral field spectrograph (Bacon et al. 1995) with an
array of 1577 square lenses, built for the William Herschel Telescope
of the Isaac Newton Group on La Palma.  One of its most notable
features is the large field of view of $33'' \times 41''$. The
instrument is the result of a collaboration between three institutes:
the Sterrewacht Leiden, the Observatoire de Lyon, and the University of
Durham. 


The data are reduced with especially developed software based on the
existing XOASIS package. An optimal extraction algorithm is used to
recover the individual spectra from the tightly packed configuration on
the CCD (Bacon et al. 2001).

The intrinsic shapes and internal dynamics of ellipticals and bulges
have long been the subject of debate. Many galactic components (e.g.\ 
halos and bars) and entire galaxies (e.g.\ giant ellipticals) are known
to have triaxial shapes. Questions that require further investigation
include: (a) What is the distribution of the intrinsic triaxial shapes?
(b) What is, at a given shape, the range in internal velocity
distributions?  (c) What is the dynamical role of massive central black
holes?  (d) What is the relation between the stellar and gaseous
kinematics and the stellar populations?

With this in mind, the SAURON project has the following goals: (a) To
provide a full census of nearby early-type galaxies along the Hubble
sequence (b) To extract complete 2D velocity fields and line strengths
for the observed galaxies (c) To combine the results with
state-of-the-art dynamical modelling and stellar population analysis

The SAURON survey has observed a representative sample of 72 nearby
ellipticals, lenticulars, and Sa bulges constructed to be as free of
biases as possible while ensuring the existence of complementary data
(e.g. HST). The galaxies are further split into ``cluster'' and
``field'' objects and populate the six $\epsilon-M_B$ planes uniformly
(de Zeeuw et al. 2002). By construction, the sample covers the full
range of environment, flattening, rotational support, nuclear cusp
slope, isophotal shape, etc.

\begin{figure}[!t]
  \includegraphics[width=\columnwidth]{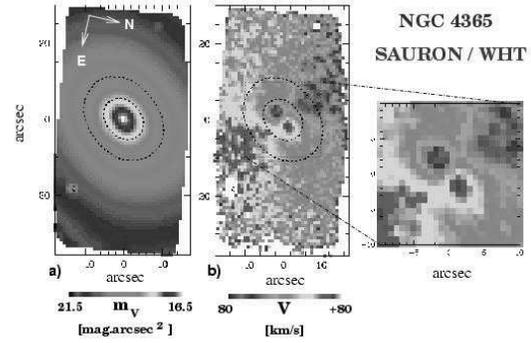}
  \caption{SAURON maps of the E3 galaxy NGC\,4365 (Davies, Kuntschner
    et al. 2001). a)~surface brightness as reconstructed from our
    data, b) mean streaming velocity $V$.
The enlarged core
    region of the velocity map shows clearly the kinematically
    decoupled core.}
  \label{fig:ngc4365}
\end{figure}

A SAURON map of NGC\,4365 with its kinematically decoupled core (KDC)
is presented in Figure~\ref{fig:ngc4365}.

\end{document}